\documentclass[aps,pra,onecolumn,nofootinbib,notitlepage,11pt,tightenlines]{revtex4-1}
\usepackage{amsmath,amssymb,amstext,braket, graphicx,color,xfrac,accents,tabulary,booktabs,xcolor,subcaption,cases}
\usepackage[unicode=true,bookmarks=true,
bookmarksnumbered=false,bookmarksopen=false,
breaklinks=false,
pdfborder={0 0 1},
backref=false,colorlinks=true]{hyperref}

\newcommand{\pdel}{\partial}

\usepackage{amsmath}  
\usepackage{amsfonts} 
\usepackage{graphicx} 

\begin{document}

\title{Restoring Invariance to the Mechanical Wave Equation across Inertial Frames}

\author{Ashmeet Singh}
\email{ashmeet@iitd.ac.in}
\affiliation{Department of Physics, Indian Institute of Technology Delhi, Hauz Khas, New Delhi 110016, India}

\author{Heather Romero Michel}
\email{romeromd@whitman.edu}
\affiliation{Physics Department, Whitman College, 345 Boyer Ave, Walla Walla, WA 99362, USA}

\hypersetup{pdftitle={Restoring Invariance to the Mechanical Wave Equation across Inertial Frames}, pdfauthor={Singh and Romero Michel}, citecolor=blue,linkcolor=blue,urlcolor=blue,citecolor=blue}


\begin{abstract}
The mechanical wave equation, derived from Newton's laws, takes its simplest form in the rest frame of the medium, and appears to lose that form entirely once expressed in the coordinates of another inertial frame --- raising the question of whether, and how, it can be written consistently for an arbitrary inertial observer. This puzzle, often glossed over in undergraduate instruction, is typically set aside by invoking the rest frame of the medium as ``preferred,'' without clarifying what this means for the principle of relativity underlying physical law. We revisit the wave equation from first principles and demonstrate that this apparent non-invariance stems from its expression in Eulerian coordinates. We present two reformulations: a manifestly invariant form using Lagrangian derivatives applied at the level of microscopic particle force laws, and a factorized Eulerian form separating left and right-moving wave solutions. In this context, we explore limitations of Galilean transformation as the low-speed limit of Lorentz transformation, and discuss potential observational connections, particularly regarding wavelength changes in the Doppler effect. Our derivation provides both conceptual clarity and pedagogical value for intermediate and advanced undergraduate physics.
\end{abstract}

\maketitle

\section{Introduction}

The wave equation is arguably one of the most important and widely applicable equations in all physics. It models spatiotemporal evolution of small disturbances under linearized force laws in diverse systems—from strings and air columns to electromagnetic fields. In its canonical form, the wave equation in one dimension reads,
\begin{equation}
    \label{eq: wave_eq_1D}
    \frac{\partial^2\psi(x, t)}{\partial t^2} = v^2 \frac{\partial^2\psi(x, t)}{\partial x^2}
\end{equation}
where $\psi(x, t)$ is the wave disturbance and $v$ is the relevant wave speed. The wave equation makes repeated appearances in the undergraduate physics curriculum in increasingly sophisticated contexts as students progress through the major. For mechanical waves, the wave equation is best thought of as an emergent phenomenon arising from applying force laws to microscopic elements of a medium, leading to a continuum, macroscopic description. Equation~(\ref{eq: wave_eq_1D}) takes its simplest form in the rest frame of the medium, where left- and right-moving disturbances propagate at equal speed $v$. It must, however, still be possible to write down a wave equation describing the same physics from the standpoint of any other inertial observer---after all, Newton's second law, from which the wave equation ultimately derives, holds equally in every inertial frame. Yet a naive attempt to do so, by substituting a change of coordinates directly into Eq.~(\ref{eq: wave_eq_1D}), does not succeed. For example, under a Galilean transformation to a frame moving at velocity $u$ relative to the medium (as we discuss in detail shortly), the canonical wave equation of Eq.~(\ref{eq: wave_eq_1D}) becomes
\begin{equation}
    \label{eq:wave_eq_1d_Sprime}
    \left[\frac{\partial^2}{\partial t'^2} - (v^2 - u^2)\frac{\partial^2}{\partial x'^2} - 2u\frac{\partial^2}{\partial x'\partial t'}\right]\psi'(x', t') = 0.
\end{equation}
This transformed equation includes a mixed-derivative convective term and a modified coefficient on the spatial second derivative, with no evident connection to the underlying Newtonian dynamics from which the wave equation emerged. This raises the central question of this paper: what is the correct way to write the wave equation for an observer in relative motion with respect to the medium, in a manner that reflects the fact that Newton's second law itself takes the same form for any such observer, and what this reveals about the motion of mechanical waves themselves.

This question, whether for electromagnetic waves or mechanical waves, is not new. The lack of form-invariance of the electromagnetic wave equation (arising from Maxwell's equations) under Galilean transformations is well known, and serves as the standard launching point for the development of special relativity in classroom treatments. But a parallel and distinct question exists for mechanical waves, which is often overlooked. In undergraduate instruction, Galilean transformations are introduced alongside the assumption of invariant accelerations, but without a careful analysis of how emergent macroscopic wave equations behave under a change of frame. It is often suggested that because mechanical waves require a medium, the rest frame of that medium provides a natural or ``preferred'' frame of reference, and that this is simply why the wave equation looks different elsewhere. This observation correctly explains why the usual expression of the wave equation (like that of Eq. \ref{eq: wave_eq_1D}) is not the appropriate one to use in arbitrary coordinates since it is formulated in terms of variables tied explicitly to the medium rest frame. However, this argument is incomplete on its own, since it leaves unanswered how the wave equation \emph{should} be written for an observer for whom the medium is in motion. Many articles, including lecture notes and teaching resources~\cite{Tegmark2005,LaussyRelativityLecture2, Ramirez2024, ng2007phy361, mitocw8.033f24, fan2024invariance} and community discussions on Physics StackExchange~\cite{Thacker2024GalileanInvariance, DavidZ2017GalileanWaveEq} quickly move past this issue, often citing the preferred-frame observation as an explanation in itself. Others propose alternate formalisms, such as modified Galilean transformations~\cite{BerishaKlinaku2023}, or directional wave solutions~\cite{Caruso2024}, or asymmetrical or modified wave equations~\cite{Musielak2023, Duran_2013}, though these approaches often sacrifice generality or lack a direct connection to Newtonian force laws.

In this work, we answer this question by recognizing that Eq.~(\ref{eq: wave_eq_1D}) is expressed in Eulerian coordinates---tracking changes at fixed points in space---rather than in a manner that reflects how Newton's second law actually acts on moving elements of a medium. The natural tool for this is the \emph{material}, or Lagrangian, derivative familiar from fluid mechanics, which follows the motion of a fluid element rather than a fixed point in space. Applying material derivatives directly to the microscopic force law governing each element of the medium, we derive a wave equation that retains a common structural form for any inertial observer, connected to Eq.~(\ref{eq: wave_eq_1D}) in the medium rest frame. We give this same result an equivalent, purely Eulerian expression by factorizing the wave operator into left- and right-moving components with frame-dependent phase velocities, which reproduces the naively transformed equation above while making its physical content, and its connection to Newtonian dynamics, transparent. In particular, we introduce a Lagrangian spatial derivative, in addition to the usual Lagrangian time derivative, which restores space-time symmetry even at the level of low-speed Lorentz transformations. Together, these results show how the motion of the medium itself, rather than any privileged status of its rest frame, accounts for the wave equation's changing appearance from frame to frame.

We carry out this analysis under both the Galilean transformation, which assumes absolute time and forms the standard backdrop of introductory mechanics, and the low-speed limit of the Lorentz transformation, which retains the relativity of simultaneity as a first-order effect while neglecting time dilation and length contraction. Along the way, this also clarifies that the Galilean transformation is not, in fact, the low-speed limit of the Lorentz transformation~\cite{Baierlein2006, Furry}, and reveals that the canonical wave equation of Eq.~(\ref{eq: wave_eq_1D}) corresponds to the special case in which left- and right-moving wave components share a common speed, true only in the medium's rest frame. The rest of this paper is structured as follows. In Section~\ref{sec:lack_of_invariance}, we show how the wave equation, taken naively into an arbitrary frame, loses its common structure under both Galilean and low-speed Lorentz transformations. In Section~\ref{sec:eulerian}, we recover this structure in a factorized Eulerian form. In Section~\ref{sec:lagrangian_reformulation}, we derive this same result from microscopic force laws using material derivatives, accounting for the relativity of simultaneity in the low-speed relativistic case. In Section~\ref{sec:discussion}, we discuss implications for pedagogy and for physical phenomena such as the Doppler shift.

\section{Lack of Manifest Invariance}\label{sec:lack_of_invariance}
\begin{figure}[h!]
    \centering
    \includegraphics[width=0.70\linewidth]{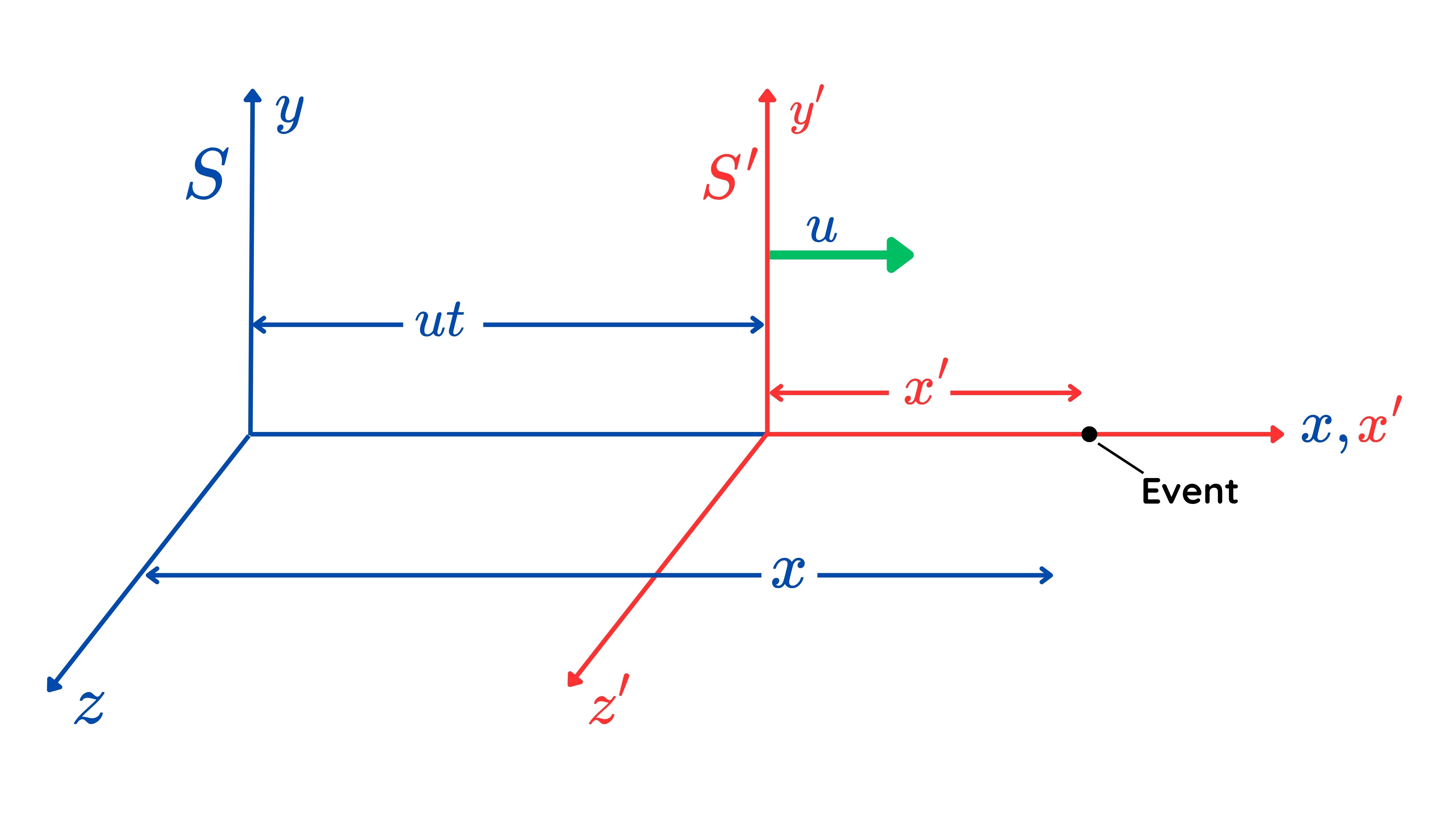}
    \caption{Schematic of inertial reference frames $S$ and $S'$ with their relative speed $u$.}
    \label{fig:IRFs}
\end{figure}

We begin by showing that the wave equation is, as expected, not invariant under either transformation between reference frames. We consider two inertial reference frames, $S$ and $S'$, moving at a relative speed $u$ along their coinciding $x$-axes. Frame $S'$ moves with speed $u$ in the $+x$ direction relative to $S$ as illustrated in Figure \ref{fig:IRFs}.

\vspace{-0.5cm}

\subsection{Galilean Relativity}
Galilean relativity asserts that the form of physical laws is the same for all inertial observers while additionally assuming time is absolute. This translates to the speed of light, or information propagation, being infinitely fast, i.e., $c = \infty$. Two inertial observers in this paradigm are connected via the Galilean transformation,
\begin{equation}
\label{eq:GT}
    x' = x - ut, \quad t' = t.
\end{equation}
The partial derivatives which enter the wave equation undergo a linear transformation via the Jacobian,
\begin{gather} \label{eq:GT_partials_matrix}
 \begin{bmatrix} \frac{\pdel}{\pdel x} \\ \\ \frac{\pdel}{\pdel t} \end{bmatrix}
 =
  \begin{bmatrix}
   \frac{\pdel x'}{\pdel x} & 
   \frac{\pdel t'}{\pdel x} \\ \\
   \frac{\pdel x'}{\pdel t} &
	\frac{\pdel t'}{\pdel t}
   \end{bmatrix} \begin{bmatrix} \frac{\pdel}{\pdel x'} \\ \\ \frac{\pdel}{\pdel t'} \end{bmatrix}
   \implies 
   \begin{bmatrix} \frac{\pdel}{\pdel x} \\ \\ \frac{\pdel}{\pdel t} \end{bmatrix}
 =
  \begin{bmatrix}
   1 &
0 \\
   -u &
	1
   \end{bmatrix} \begin{bmatrix} \frac{\pdel}{\pdel x'} \\ \\ \frac{\pdel}{\pdel t'} \end{bmatrix}
   ,
\end{gather}
which when substituted into the wave equation of Eq. (\ref{eq: wave_eq_1D}) gives us the following transformed form,
\begin{equation}
\label{eq:wave_eq_1d_Sprime}
\left[\frac{\partial^2}{\partial t'^2} - (v^2-u^2)\frac{\partial^2}{\partial x'^2} - 2u\frac{\partial}{\partial t'\partial x'}\right]\psi'(x',t')= 0 \: ,
\end{equation}
where $\psi'(x',t')$ denotes the same physical disturbance as $\psi(x,t)$, now expressed as a function of the $S'$ coordinates.\footnote{By construction, $\psi$ and $\psi'$ describe one and the same physical quantity, so their values agree at corresponding spacetime points, $\psi'(x',t') = \psi(x,t)$, even though, as Eq.~(\ref{eq:wave_eq_1d_Sprime}) makes clear, their functional dependence on the respective coordinates is not the same.} This transformed equation is structurally different from the original: it is not simply a more complicated expression of the same differential operator, but contains an additional mixed derivative term and a modified coefficient on the spatial second derivative that are absent from Eq.~(\ref{eq: wave_eq_1D}), with no obvious relation to the expected velocity addition $v \pm u$. By this we mean a lack of manifest form-invariance at the level of the differential operator itself, rather than at the level of its solutions. It is neither surprising nor in tension with the principle of relativity that particular solutions can take a simpler form in one frame than another—the medium rest frame is itself an example, where left- and right-moving waves propagate at equal speed. What is notable here is instead that the governing equation, prior to any choice of solution, takes a different functional form once boosted. This apparent frame dependence suggests, misleadingly, that the wave equation privileges the rest frame of the medium---contradicting the principle of relativity, which requires that the equations of motion, not merely their solutions, retain the same form in every inertial frame. This issue has been noted in prior works on the topic~\cite{BerishaKlinaku2023,Caruso2024}, although without a satisfactory derivation from microscopic force laws or a pedagogically accessible resolution.


\subsection{Low-Speed Limit of Special Relativity (Lorentz Transformation)} \label{sec:LLT}
Similarly to the Galilean relativity case, the wave equation does not remain invariant under Lorentz transformation, including in its low speed limit, which is the focus of our work. We aim to capture the physically relevant first-order relativistic effects that persist even in non-relativistic regimes where $u/c, v/c \ll 1$. The standard Lorentz transformation between inertial frames $S$ and $S'$, with $S'$ moving at speed $u$ in the $+x$ direction, is given by:
\begin{equation}
\label{eq:LT}
    x' = \gamma(x - u t), \;\;\; t' = \gamma(t - \frac{u}{c^{2}} x) \: ,
\end{equation}
where $\gamma = (1 - (u/c)^2)^{-1/2}$. By focusing only on first-order effects, \textit{i.e.} by retaining terms linear in $u/c$, we can expand $\gamma$ to obtain $\gamma \approx 1 + \mathcal{O}((u/c)^2)$. This expansion shows that time dilation and length contraction are second-order effects and negligible at low speeds. However, the $ux/c^2$ term in the time transformation is first order in $u/c$, and thus remains significant. It captures the relativity of simultaneity, the fact that observers in relative motion disagree on whether spatially separated events occur at the same time. Thus, Lorentz transformation does not reduce to its Galilean counterpart when terms linear in $u/c$ are preserved. In fact, the two differ fundamentally in how they treat time. This underappreciated but highly important point is emphasized in Furry’s classic 1955 article~\cite{Furry}, or in more contemporary works by Baierlein~\cite{Baierlein2006} and Chen et al.~\cite{Chen2022}, where it is shown that accounting for only first-order terms in the Lorentz transformation leads to a loss of simultaneity. Time is, therefore, not absolute even at low-speeds. The appearance of relativity of simultaneity as a relativistic effect does not depend on the use of large relative speed of the two reference frames, but only on the choice of spacetime events\footnote{As long as $u \neq 0$, it is always possible to find a correspondingly large value of $x$ that makes the second term in the temporal transformation non-negligible.} and the precision of measured time intervals. The correct low-speed approximation to the Lorentz transformation is thus:
\begin{equation}
\label{eq:lowlorentz}
x' = x - ut, \quad t' = t - \frac{ux}{c^2}.
\end{equation}
This transformation is further confirmed by the group structure of Lorentz transformation. Low-speed Lorentz transformations of Eq. (\ref{eq:lowlorentz}) can be composed to produce finite Lorentz transformations of Eq. (\ref{eq:LT}), whereas no amount of infinitesimal or finite Galilean transformations of Eq. (\ref{eq:GT}) can be composed to generate a Lorentz transformation \cite{Furry}. This underscores the point that a Galilean transformation does not correspond to the limit $u/c \ll 1$. There are two distinct ways in which low-speed Lorentz transformation of Eq. (\ref{eq:LT}) reduces to its Galilean counterpart of Eq. (\ref{eq:GT}): (i.) in the non-physical limit $c \to \infty$ where information propagates instantaneously, and/or (ii.) if the spatial interval $\Delta x \rightarrow 0$ (which reduces to $x$ if one of the points is the origin) making the relativity of simultaneity vanish. 
Under the low-speed Lorentz transformation of Eq. (\ref{eq:lowlorentz}), the partial derivatives that enter the wave equation undergo the following transformation,
\begin{gather} \label{eq:LLT_partials_matrix}
 \begin{bmatrix} \frac{\pdel}{\pdel x} \\ \\ \frac{\pdel}{\pdel t} \end{bmatrix}
 =
  \begin{bmatrix}
   \frac{\pdel x'}{\pdel x} & 
   \frac{\pdel t'}{\pdel x} \\ \\
   \frac{\pdel x'}{\pdel t} &
	\frac{\pdel t'}{\pdel t}
   \end{bmatrix} \begin{bmatrix} \frac{\pdel}{\pdel x'} \\ \\ \frac{\pdel}{\pdel t'} \end{bmatrix}
   \implies 
   \begin{bmatrix} \frac{\pdel}{\pdel x} \\ \\ \frac{\pdel}{\pdel t} \end{bmatrix}
 =
  \begin{bmatrix}
   1 &
   -\frac{u}{c^{2}} \\
   -u &
	1
   \end{bmatrix} \begin{bmatrix} \frac{\pdel}{\pdel x'} \\ \\ \frac{\pdel}{\pdel t'} \end{bmatrix}
   ,
\end{gather}
which when substituted into the wave equation (\ref{eq: wave_eq_1D}) gives us the following transformed form,
\begin{equation}
\label{eq:lowlorentz_wave_transformed}
    \left[\frac{\partial^2}{\partial t'^2} -\frac{(v^2-u^2)}{\left(1-\frac{v^2u^2}{c^4}\right)}\frac{\partial^2}{\partial x'^2} - 2u\frac{\left(1-\frac{v^2}{c^2}\right)}{\left(1-\frac{v^2u^2}{c^4}\right)}\frac{\partial^2}{\partial x'\partial t'}\right]\psi'(x',t')=0 \: .
\end{equation}
It is important to note that although the transformation in Eq. (\ref{eq:LT}) is truncated at first order in $u/c$, the second derivatives in the wave equation generate terms of order $(u/c)^2$ upon substitution. These arise algebraically from the structure of the differential operator, not from higher-order terms in the transformation itself. To maintain consistency with the order of approximation, we systematically discard such $(u/c)^2$ terms while retaining all terms linear in $u/c$, under which Eq. (\ref{eq:lowlorentz_wave_transformed}) becomes:
\begin{equation}
\label{eq:lowlorentz_wave_transformed_linear}
    \left[\frac{\partial^2}{\partial t'^2} -(v^2-u^2)\frac{\partial^2}{\partial x'^2} - 2u\left(1-\frac{v^2}{c^2}\right)\frac{\partial^2}{\partial x'\partial t'}\right]\psi'(x',t')=0 \: .
\end{equation}
In our analysis, we perform an expansion only in the boost parameter $u/c$, while treating the wave speed $v$ as a fixed material parameter determined by the properties of the medium. The quantity $v$ characterizes the intrinsic propagation speed of disturbances in the medium and is independent of the boost velocity $u$. As a result, combinations such as $uv^2/c^2$ are linear in the expansion parameter $u/c$ and are consistently retained, while terms of order $u^2/c^2$ and higher are neglected. Terms such as $u^2$ that appear multiplying the second spatial derivative arise already at the level of Galilean kinematics (and are not necessarily small since they do not involve powers of $u/c$ but simply $u$) and are therefore retained exactly, as they are not associated with relativistic corrections. 

%

\section{Eulerian Invariance: A Separable Form} \label{sec:eulerian}
In systems involving fluid flow, wave propagation, or continuum mechanics in general, two complementary frameworks are commonly employed to describe the dynamics: the Eulerian and Lagrangian perspectives. In the Eulerian framework, one observes how physical quantities such as pressure, displacement, or velocity evolve at fixed points in space or time. The partial derivatives which appear in the wave equation, such as that of Eqs. (\ref{eq: wave_eq_1D}), (\ref{eq:wave_eq_1d_Sprime}), (\ref{eq:lowlorentz_wave_transformed}) are Eulerian derivatives since they track changes at a given space or time point. The Lagrangian framework, by contrast, follows individual fluid elements or material particles as they move through space, tracking the evolution of quantities along their trajectories. The invariance of physical laws should be manifest in both frameworks, which we now show in the context of the mechanical wave equation. We tackle the Lagrangian picture in Section \ref{sec:lagrangian_reformulation} where we derive an invariant form directly from Newton’s second law applied to infinitesimal elements of the medium, using material derivatives that respect the motion of those elements. While the Lagrangian picture will offer a more satisfying construction, we first work in the Eulerian picture to demonstrate that invariance is recovered when the wave equation is written in a factorized form that separates left- and right-moving modes with frame-dependent phase velocities. To begin, we note that in the rest frame of the medium, both right- and left-moving plane wave solutions travel at the same speed $v$, and the d’Alembertian in the canonical wave equation of Eq.~\eqref{eq: wave_eq_1D} can be factorized to yield two first-order partial differential expressions
\begin{equation}
    \label{eq:conjugate_RF}
    \left[\left(\frac{\partial}{\partial t} + v\frac{\partial}{\partial x}\right)\left(\frac{\partial}{\partial t} - v\frac{\partial}{\partial x}\right)\right]\psi(x,t) = 0.
\end{equation}
This form makes explicit that the general solution is composed of right- and left-moving waves with velocities $\pm v$. However, in a frame moving at speed $u$ relative to the medium rest frame, the observed phase velocities of these waves from this frame are no longer symmetric; they must transform according to the appropriate velocity addition rules. Cognizant of this asymmetry, we propose writing a generalized form of the wave equation in an arbitrary inertial frame, in which the right- and left-moving waves have (possibly different) phase speeds $v_r$ and $v_l$, respectively,
\begin{equation}
    \label{eq:conjugate_arbit}
    \left[\left(\frac{\partial}{\partial t} + v_{r}\frac{\partial}{\partial x}\right)\left(\frac{\partial}{\partial t} - v_{l}\frac{\partial}{\partial x}\right)\right]\psi(x,t) = 0.
\end{equation}
We refer to this as the \textit{factorized form} of the wave equation. We note that the speeds $v_r$ and $v_l$ are not independent parameters. Instead, they depend on $v$, the wave speed in the medium rest frame, and $u$, the relative speed of the frames, as we show below. 
Importantly, this formulation of Eq. (\ref{eq:conjugate_arbit}) preserves the same structure in every inertial reference frame, even though the specific values of $v_r$ and $v_l$ may vary depending on the frame (and in turn depend on $v$ and $u$). We show in Section \ref{sec:lagrangian_reformulation} that this result can be connected with the Lagrangian framework and maintains its form across all inertial frames, whether connected by Galilean or low-speed Lorentz transformations. Expanding this product yields:
\begin{equation}
    \label{eq:conjugate_arbit2}
    \left[\frac{\partial^2}{\partial t^2} - (v_{r}v_{l})\frac{\partial^2}{\partial x^2} + (v_{r} - v_{l})\frac{\partial^2}{\partial x \partial t}\right]\psi(x,t) = 0.
\end{equation}


In an arbitrary frame moving with speed $u$ relative to the medium, Galilean velocity addition gives $v_r = v - u$, and $v_l = v + u$, 
leading to $v_r v_l = v^2 - u^2,$ and $v_r - v_l = -2u$, which recovers the Galilean transformed wave equation in Eq.~\eqref{eq:wave_eq_1d_Sprime}. An interesting re-writing of the Galilean transfomed wave equation of Eq. (\ref{eq:wave_eq_1d_Sprime})  offers an important foresight into the resolution to restore invariance to the wave equation,
\begin{equation}
\left[\bigg(\frac{\partial}{\partial t'} - u \frac{\partial}{\partial x'} \bigg)^{2} - v^{2} \frac{\partial^{2}}{\partial x'^{2}} \right] \psi(x',t') = 0 \: .
\end{equation}
One recognizes $(\partial_{t'} - u\partial_{x'	})$ as the \emph{Lagrangian} (or material) derivative, which suggests using a microscopic, particle-based force approach, something we will explore in detail in Section \ref{sec:lagrangian_reformulation}.

Under a low-speed Lorentz transformation as in Eq.~\eqref{eq:lowlorentz}, the phase velocities transform as:
\begin{equation}
    v_{r} = \frac{v - u}{1 - uv/c^{2}} = (v - u) + \frac{uv^{2}}{c^{2}} +\mathcal{O}\left(\frac{u}{c}\right)^{2}, \quad  v_{l} = \frac{v + u}{1 + uv/c^{2}}  = (v+ u) -\frac{uv^{2}}{c^{2}} +\mathcal{O}\left(\frac{u}{c}\right)^{2} ,
\end{equation}
where the second equality in each equation helps identify terms linear in $u/c$ that are relevant to our analysis. This then gives to \emph{linear order in $u/c$},
\begin{equation}
    v_{r}v_{l} \approx v^{2} - u^{2}, \quad v_{r} - v_{l} \approx -2u\left(1 - v^{2}/c^{2}\right),
\end{equation}
which precisely reproduces the low-speed Lorentz transform of the wave equation  \eqref{eq:lowlorentz_wave_transformed_linear}. This analysis demonstrates that the factorized form of the wave equation, which explicitly distinguishes left- and right-moving components and incorporates appropriate velocity addition, retains the same structural form across all inertial frames. We will now derive this form of Eq. (\ref{eq:conjugate_arbit2}) from first-principle microscopic force laws. 


\section{Lagrangian Invariance from Microscopic Force Laws}\label{sec:lagrangian_reformulation}

\begin{figure}[h!]
    \centering
    \includegraphics[width=0.75\linewidth]{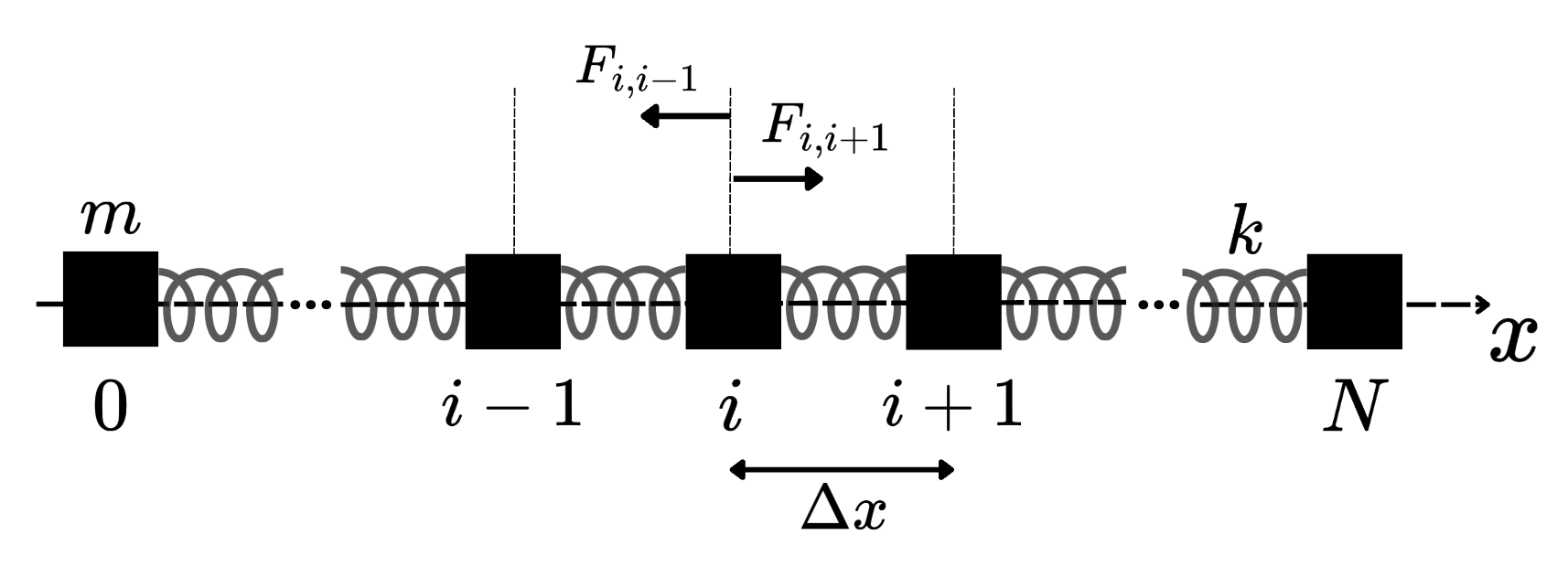}
    \caption{Schematic of 1-D chain of nearest neighbor coupled oscillators with linear spring-like restoring forces}
    \label{fig:1D_chain}
\end{figure}

To connect the macroscopic wave equation with its microscopic origins, we now shift to the Lagrangian framework, where we follow individual fluid elements or particles as they move through space and time. This shift is essential because Newton’s second law, $ \sum\vec{F} = m\vec{a}$ , applies to material particles and governs their accelerations along their trajectories. In both Galilean and low-speed Lorentz frameworks, the appropriate force law is Newton’s second law in its familiar form. Although this is clearly valid in the Galilean case, its use in the low-speed relativistic regime requires justification. Starting from the fully relativistic expression for the net force\footnote{One can alternatively also use a relativistically covariant Four-Force vector, $F^{\mu} = dp^{\mu}/d\tau$, the spatial part of which reduces to Eq. (\ref{eq:rel_force}) in the $v_{p}/c \ll 1$ limit where $dt/d\tau = \gamma \approx 1$. See \cite{Jefimenko1996}, for example.} on a particle of mass \( m \) and velocity \( \vec{v}_{p} \) (note this particle velocity $\vec{v}_{p}$ is in general distinct from rest frame wave speed $v$ and Lorentz boost speed $u$),
\begin{equation}\label{eq:rel_force}
    \vec{F} = \frac{d\vec{p}}{dt} = \frac{d}{dt}\left(\frac{m \vec{v}_{p}}{\sqrt{1 - v^{2}_{p}/c^{2}}}\right) = \gamma_{p} m \vec{a} + \gamma^{3}_{p} m \frac{(\vec{v}_{p}\cdot\vec{a}) \vec{v}_{p}}{c^{2}} \: ,
\end{equation}
where $\vec{a} = d\vec{v}_{p}/dt$ is the particle's acceleration and $\gamma_{p}$ is the Lorentz factor relevant for the particle speed $v_p$. We see that in one spatial dimension this simplifies to
\begin{equation}
    F = \gamma_{p}^{3} m a  \: .
\end{equation}
This expression for the force law is indeed covariant under a full Lorentz transformation, $F' = \gamma'^{3}_{{p}} m a'$ defined on coordinates $(x',t')$ of Eq. (\ref{eq:LT}), with the following transformed quantities,
\begin{equation}
v_{p}' = \frac{v_{p} - u}{1 - \frac{u v_{p}}{c^{2}}} \: , \quad a' = \frac{a}{\gamma^{3}_{u} \left(1 - \frac{uv_{p}}{c^{2}}\right)^{3}} \: , \quad \gamma'_{p} = \gamma_{p}\gamma_{u}\left(1 - \frac{uv_{p}}{c^{2}}\right) \: .
\end{equation}
We wish to study the form and invariance of this force law under two approximations: (1) the Lorentz boost is linear in  $u/c$, and (2) the speed of the particle is small compared to $c$, that is, by retaining only linear order in $v_{p}/c$. This second condition is tied with small amplitude, linear wave solutions we are dealing with, where each oscillating particle (such as in a chain of coupled oscillators) has a maximum speed $v_{p} \lesssim A \omega$ where $A$ is the amplitude of the oscillation and $\omega$ is the angular speed --- small amplitudes of the wave solutions require particle speeds to also be small. Hence, as a consequence of $u/c \ll 1$ and $v_{p}/c \ll 1$, we neglect terms of order $(u/c)^{2}, (v_{p}/c)^{2}, uv_{p}/c^{2}$ and higher. As \( \gamma_{p} = 1 + \mathcal{O}(v_{p}^{2}/c^{2}) \) and $\gamma_{u} = 1 + \mathcal{O}(u/c)^{2}$, this shows that the force law in this regime is $F = ma$ and is invariant under first-order Lorentz transformation for small amplitude motion. We mention at this stage that relativistic corrections at order $(v_{p}/c)^{2}$ will make the emergent wave equation non-linear, something we will not focus on in this article. Before we apply Newton's laws, it is important to ask which kind of derivatives should one use? Is the standard $\pdel^{2}/\pdel t^{2}$ sufficient to capture the acceleration of a specific particle? Or do we need a different approach?
In what follows, we focus on a one-dimensional chain of \( N \gg 1 \) identical masses (each of mass $m$) connected by linear springs, a standard model for longitudinal waves such as sound propagation in solids. Each spring has natural length \( \Delta x \) and spring constant \( k \) as depicted in Fig. \ref{fig:1D_chain}. The continuum limit of this setup will yield the macroscopic wave equation as an emergent phenomenon, allowing us to derive its invariant form directly from Newton's laws.
For clarity and completeness, we analyze the microscopic dynamics in two frames: the rest frame $S$ of the medium, and an {arbitrary observer frame} $S'$ moving with velocity \( u \) relative to the medium as shown in Fig (\ref{fig:IRFs}). We will not do any frame transformations between $S$ and $S'$, but rather {derive} the wave equation in the arbitrary frame $S'$ from first principles. Thus, whenever we will now address the wave equation in the primed coordinates, it will be taken to represent the dynamics in that frame\footnote{The case of the rest frame $S$ can also be recovered by simply setting $u=0$ in the $S'$ derivation.}. In the rest frame \( S \) of the medium, the displacement of the \( i \)-th particle (with equilibrium location at $x$) is described by a function \( \psi(x, t) \), and the wave propagates in either the \( +x \) or \( -x \) direction with constant speed \( v \). This temporal variation is captured by the Eulerian second time derivative \( \partial^2 \psi / \partial t^2 \), which measures the acceleration at a fixed point in space. From the viewpoint of~$S'$, the phase velocity of the wave is altered by the relative motion. Consequently, at a fixed position~$x'$, the observed displacement varies in time not only due to intrinsic oscillations of the particles but also because, in frame~$S'$, different particles move past the fixed~$x'$. The corresponding Eulerian acceleration in this frame is \( \partial^2 \psi'/\partial t'^2 \), and generally we find that they do not agree across frames, $\frac{\partial^2 \psi(x, t)}{\partial t^2} \neq \frac{\partial^2 \psi'(x', t')}{\partial t'^2}$.
In fact, in the special case where the right-moving wave appears stationary in \( S' \) (i.e., when \( u = v \)), the Eulerian time derivative in \( S' \) vanishes entirely—even though individual particles are clearly accelerating. This illustrates the scope of the Eulerian viewpoint: it does not track individual elements of the medium and thus fails to represent the invariant dynamics governed by Newton's laws.

The appropriate resolution is to adopt the Lagrangian (also commonly referred to as the material, or convective) derivative \cite{french1971vibrations}, which follows the motion of a specific mass element in the system. We derive its explicit form directly from the motion of a single discrete particle. In the rest frame $S$, the $i$-th particle carries a fixed equilibrium label $x_i = i\Delta x$ and has instantaneous position $X_i(t) = x_i + \eta_i(t)$, with $\eta_i(t) \equiv \psi(x_i, t)$ its displacement. Since Galilean transformations leave time and length intervals invariant, the same particle's position in the frame of observer $S'$ is $X_i'(t') = X_i(t) - ut' = (x_i - ut') + \eta_i(t')$. Here, we can identify $x_i - ut'$ as the particle's drifting equilibrium position in $S'$ since the entire lattice moves at velocity $-u$ as seen by $S'$ (since $S'$ moves at $+u$ relative to the medium). Newton's second law for this single particle is then an ordinary second derivative along its trajectory,
\begin{equation}
    \label{eq:discrete_newton}
    m\,\frac{d^2 X_i'(t')}{dt'^2} = F_i(t') \quad \Longrightarrow \quad m\,\frac{d^2 \eta_i(t')}{dt'^2} = F_i(t') \: ,
\end{equation}
where the second form follows because the equilibrium drift term $x_i - ut'$ is linear in $t'$ and does not contribute to the acceleration. At this stage $\eta_i(t')$ is a function of the single variable $t'$ for each fixed label $i$.

We now pass to a continuum description. Because we are interested in disturbances whose wavelength is long compared to the lattice spacing $\Delta x$, the displacements $\eta_i(t')$ may be regarded as samples of a smooth field, $\eta_i(t') \equiv \psi'(x_i'(t'), t')$, evaluated at each particle's drifting equilibrium coordinate $x_i'(t') = x_i - ut'$. This is the first of two continuum limits we take: the present one serves only to render $\psi'$ a differentiable function of $x'$, whereas the nearest-neighbor force law of Eq.~(\ref{eq:force_base}) is kept in exact discrete form and Taylor-expanded in its own $\Delta x \to 0$ limit later, where the second spatial derivative and the emergent wave speed $v$ arise (Eqs.~(\ref{eq:Taylorseries1})--(\ref{eq:galilean_wave_lagrangian})). Since the label $x_i$ is a fixed coordinate, the difference between displacements of neighbors, divided by the equilibrium separation, in the continuum limit, is precisely the spatial derivative of the field,
\begin{equation}
    \label{eq:partial_x_from_lattice}
    \frac{\partial \psi'}{\partial x'}\bigg|_{x_i'(t'),\,t'} = \lim_{\Delta x \to 0} \frac{\eta_{i+1}(t') - \eta_i(t')}{\Delta x} \: .
\end{equation}

Differentiating the composite function $\eta_i(t') = \psi'(x_i'(t'), t')$ by the chain rule, and using Eq.~(\ref{eq:partial_x_from_lattice}) for its spatial part, gives
\begin{equation}
    \label{eq:chain_rule_discrete}
    \frac{d\eta_i}{dt'} = \frac{d}{dt'}\Big[\psi'(x_i'(t'), t')\Big]
    = \frac{\partial \psi'}{\partial t'}\bigg|_{x_i'(t'),\,t'}
    \: + \: \frac{dx_i'}{dt'}\,\frac{\partial \psi'}{\partial x'}\bigg|_{x_i'(t'),\,t'} \: .
\end{equation}
Since $dx_i'/dt' = -u$ is the velocity of the drifting lattice in $S'$, this reduces to
\begin{equation}
    \label{eq:chain_rule_discrete_operator}
    \frac{d\eta_i}{dt'} = \left(\frac{\partial}{\partial t'} - u \frac{\partial}{\partial x'}\right)\psi'(x',t')\bigg|_{x' = x_i'(t')} \: .
\end{equation}
This identifies the bracketed operator as the Lagrangian time derivative,
\begin{equation}
    \label{eq:Lagrangian_der}
    \frac{D}{Dt'} \equiv \frac{\partial}{\partial t'} - u \frac{\partial}{\partial x'} \: ,
\end{equation}
obtained directly from tracking a single labeled particle through a slowly-varying displacement field. This operator tracks the time rate of change a specific point on the lattice chain, the operator relevant for obtaining the particle's acceleration when applying \( F = ma \). We note that Eq. (\ref{eq:Lagrangian_der}) holds for any $u$, not necessarily small. While Eq. (\ref{eq:Lagrangian_der}) resembles the time derivative transformations under both Galilean and low-speed Lorentz transformations (see Eqs. \ref{eq:GT_partials_matrix} and \ref{eq:LLT_partials_matrix}), however, here it is connecting Eulerian and Lagrangian derivatives in the \emph{same} frame $S'$. In the rest frame of the medium, the Eulerian and Lagrangian derivatives coincide since $u = 0$.

We now confirm that this operator, obtained above from a physical argument about
the particle's velocity in $S'$, is consistent with directly transforming the
rest-frame Lagrangian derivative $D/Dt$ under the Galilean transformation
itself, i.e., we compare $D/Dt$ on $\psi(x,t)$ evaluated in $S$ against
$D/Dt'$ acting on $\psi'(x',t')$ of Eq.~\eqref{eq:Lagrangian_der} evaluated in $S'$, and
show the two agree once expressed in a common set of coordinates. We first
isolate, from the Jacobian of Eq.~\eqref{eq:GT_partials_matrix}, the
transformation of the time derivative alone: for any differentiable function
of $(x',t')$,
\begin{equation}
    \label{eq:GT_time_op}
    \frac{\partial}{\partial t}\bigg|_{x} \: = \: \frac{\partial}{\partial t'} - u \frac{\partial}{\partial x'} \: .
\end{equation}
Since the physical disturbance takes the same value at a given spacetime point
in either frame, and $\psi'(x',t')$ is \emph{defined} by rewriting the
arguments of $\psi$ via Eq.~\eqref{eq:GT}, we have the chain of equalities
\begin{equation}
    \label{GT equivalent functions}
    \psi(x,t) \: \overset{\mathrm{Eq. } \eqref{eq:GT}}{=} \: \psi(x'+ut',t') \: \overset{\text{def.}}{=} \: \psi'(x', t') \: \overset{\mathrm{Eq. } \eqref{eq:GT}}{=} \: \psi'(x - ut, t) \: ,
\end{equation}
where the first and third equalities substitute the forward and inverse
Galilean transformations, respectively, and the second is simply the
definition of $\psi'$. Applying $\partial/\partial t|_x$ to the first form in
Eq.~\eqref{GT equivalent functions}, and using Eq.~\eqref{eq:GT_time_op} on
the second, gives
\begin{equation}
    \label{Lagrangian Derivative S = S' step 1}
    \underbrace{\frac{D}{Dt}\psi(x,t) = \frac{\partial}{\partial t}\psi(x,t)}_{\text{Rest Frame } S \: (u=0)}
    \: = \:
    \left[\frac{\partial}{\partial t'} - u \frac{\partial}{\partial x'}\right]\psi(x'+ut',t')
    \: \overset{\eqref{eq:Lagrangian_der}}{=} \:
    \frac{D}{Dt'}\psi'(x',t') \: ,
\end{equation}
where the final equality recognizes the bracketed operator, acting on
$\psi'(x',t')$, as precisely $D/Dt'$ of Eq.~\eqref{eq:Lagrangian_der}. This
confirms that the Lagrangian derivative yields the same velocity, and by
extension, acceleration, in both frames, as expected from a kinematical
quantity invariant under the principle of relativity. Thus, crucially, when writing down accelerations of individual mass elements in $F = ma$, we need to use Lagrangian derivatives in the observer's frame, for example, for the $i$-th particle, $a'_i = D^{2}\psi'_i/Dt'^{2}$, which is obtained by applying the Lagrangian time derivative of Eq. (\ref{eq:Lagrangian_der}), twice,
\begin{equation}
\label{eq:Lagrangian_acc}
    \frac{D^{2}}{Dt'^{2}} \: = \: \frac{\pdel^{2}}{\pdel t'^{2}} - 2u\frac{\pdel^{2}}{\pdel x' \pdel t' } + u^{2} \frac{\pdel^{2}}{\pdel x'^{2}} 
\end{equation}
The other non-trivial part of the force law is the force itself. The force on the $i$-th mass in the 1D chain of coupled oscillators as shown in Fig. (\ref{fig:1D_chain}) will be a combination of restoring forces from its nearest neighbors~\cite{french1971vibrations},
\begin{equation}
\label{eq:force_base}
    F_{i} = k \left(\psi_{i+1} - 2\psi_{i} + \psi_{i-1}\right) \: ,
\end{equation}
where $k$ is the spring constant. We deliberately leave out spacetime points on which each of the $\psi_{j}$ terms are evaluated since we will treat them carefully in the upcoming subsections separately dealing with Galilean relativity (where speed of information propagation $c = \infty$) and low-speed special relativity (where simultaneity is relative due to finite speed of light or equivalently, of information propagation, $c < \infty$).

\subsection{Galilean Relativity ($c = \infty$)}
In an arbitrary observer's frame, $S'$ (not necessarily rest frame of medium), the motion of individual particles along the longitudinal chain of coupled oscillators is governed by Newton's second law, \( F = ma \), applied using Lagrangian derivatives to track the acceleration of each moving particle. For the $i$-th particle located at position $x'$ in frame $S'$, the equation of motion becomes
\begin{equation}
\label{eq:Newton_wave_Galilean}
    m\frac{D^{2}}{Dt'^{2}} \psi'(x',t') \: = F(x',t') \: = \: k \left[\psi'(x' + \Delta x,t') - 2\psi'(x' ,t') + \psi'(x' - \Delta x,t')\right] \: ,
\end{equation}
where we have used the nearest-neighbor interaction from Eq. (\ref{eq:force_base}). \( \psi'(x', t') \) is the displacement from equilibrium and \( \Delta x \) is the equilibrium spacing between particles. Note that under Galilean relativity, there is no relativistic length contraction, so \( \Delta x \) is invariant across frames.

A key feature here is that the force on the $i$-th particle depends on the positions of its nearest neighbors \emph{at the same time $t'$} since information propagates instantaneously ($c=\infty$) in Galilean relativity. This will not be the case with low-speed special relativistic treatment of the next subsection since light takes a finite time to propagate information.

To transition to a continuous description, we first perform a Taylor expansion of the displacements around $x'$:
\begin{equation}
\label{eq:Taylorseries1}
    \psi'(x' \pm \Delta x, t') \: = \: \psi'(x' , t')  \: \pm \: \Delta x \frac{\pdel \psi'}{\pdel x'}\bigg|_{(x',t')} \: + \: \frac{1}{2}\left(\pm \Delta x\right)^{2} \frac{\pdel^{2} \psi'}{\pdel x'^{2}}\bigg|_{(x',t')} \: + \: \mathcal{O}(\Delta x)^{3} \: .
\end{equation}
Substituting this expansion into Eq. (\ref{eq:Newton_wave_Galilean}), we notice that the linear terms cancel out. Taking the continuum limit \( \Delta x \rightarrow 0 \), we use the identification \( m = \rho \Delta x \), where \( \rho \) is the linear mass density, and define a wave speed \( v = \sqrt{k \Delta x / \rho} \equiv \sqrt{B/\rho} \) defined relative to the medium rest frame, where \( B \) is the bulk modulus of the medium, we get
\begin{equation}\label{eq:galilean_wave_lagrangian}
    \frac{D^2}{Dt'^2} \psi'(x', t') = v^2 \frac{\partial^2 \psi'}{\partial x'^2} \: .
\end{equation}
We stress that $v$ is the emergent wave speed, and is dependent on the material properties, such as its elasticity and density, and is a fixed physical scale. Expressing the Lagrangian acceleration in terms of partial (Eulerian) derivatives via Eq. (\ref{eq:Lagrangian_acc}), recovers the ``transformed'' form of the wave equation under Galilean transformations, as previously derived in Eq. (\ref{eq:wave_eq_1d_Sprime}),
\begin{equation}
    \left[\frac{\partial^2}{\partial t'^2} - (v^2 - u^2) \frac{\partial^2}{\partial x'^2} - 2u \frac{\partial^2}{\partial t' \partial x'}\right] \psi'(x', t') = 0 \: ,
\end{equation}
but interpreted here as being the wave equation derived in an arbitrary inertial frame emerging from invariant Newton's laws. It matches the general invariant form of Eq. (\ref{eq:conjugate_arbit2}) when the left- and right-moving wave components are assigned Galilean-shifted phase velocities \( v_l = v + u \) and \( v_r = v - u \). Finally, as a special case, in the rest frame of the medium, where \( u = 0 \), we recover the canonical wave equation of Eq. (\ref{eq: wave_eq_1D}).

\subsection{Low-Speed Limit of Special Relativity ($c < \infty \: \mathrm{and} \: u/c \ll 1$)} 

The analysis in the low-speed limit of special relativity is more subtle than in the Galilean case due to the finite speed of information propagation. In this regime, simultaneity is no longer absolute, and we must carefully account for signal delays when modeling forces between interacting particles. To establish the framework, we begin our analysis in the rest frame $S$ of the medium. At time $t$, the $i$-th particle located at $x$ experiences restoring forces from its nearest neighbors, but those forces are due to their \emph{retarded} positions—i.e., their displacements at an earlier time $t_* = t - \Delta t_0$, where $\Delta t_0 = \Delta x / c$ is the time it takes for information to propagate a distance $\Delta x$ at speed $c$. We take $\Delta x$ to be small since we are working with small amplitudes relevant for the linear wave equation. The force law in this frame thus takes the form (Lagrangian and Eulerian derivatives are identical in medium rest frame $S$)
\begin{equation}
    \: m\frac{\partial^{2} \psi(x,t)}{\partial t^{2}} = k \left[\psi(x + \Delta x, t_*) - 2\psi(x,t) + \psi(x - \Delta x, t_*)\right] \: .
\end{equation}
This motivates the definition of a \emph{spacetime-local} second spatial derivative that includes retardation effects (we do not use separate notation to indicate this to avoid clutter)
\begin{equation}
\label{eq:spacetime_local_secondx}
    \frac{\partial^2 \psi(x,t)}{\partial x^2} \equiv \lim_{\Delta x \to 0} \left[\frac{\psi(x+\Delta x,t_*) - 2\psi(x,t) + \psi(x-\Delta x,t_*)}{\Delta x^2} \right] \: .
\end{equation}
More concretely, this captures the curvature of the wave form, in a local form, both spatially and temporally (\textit{i.e.} at a given spacetime location $(x,t)$ accounting for finite speed of information propagation). With this redefinition, the standard wave equation remains valid in its canonical form even when considering low-speed limit of special relativity, $\frac{\partial^2 \psi(x,t)}{\partial t^2} = v^2 \frac{\partial^2 \psi(x,t)}{\partial x^2}$, where $v = \sqrt{B/\rho}$ is the wave speed, as before.

Now consider the same system viewed from the inertial frame $S'$ of an arbitrary observer moving with velocity $u$ relative to the medium. Two key changes must now be made in the analysis. First, we must replace Eulerian time derivatives with Lagrangian ones, since we are now tracking fluid elements in motion. Second, the restoring forces must account for the relativity of simultaneity, that is, the retarded positions of neighboring particles are not recorded at equal times in $S'$. Specifically, as the particles move along the $-x'$ direction with a speed of $u$ relative to the observer $S'$, the $i$-th particle at time $t'$ will respond to the position of the $(i \pm 1)$-th particle at $t' - \Delta t_{\pm}$ where,
\begin{equation}
    \Delta t_\pm = \frac{\Delta x}{c \mp u} \approx \Delta t_0 \pm \frac{u \Delta x}{c^2} \: ,
\end{equation}
where in the second (approximate) equality, we have retained leading order terms. We identify $\Delta x = \Delta x'$ since length contraction effects are second-order in $u/c$ and we are considering only first-order relativistic effects. In effect, the particle receives news from its left neighbor slightly earlier than from its right — this is relativity of simultaneity in action. The force law $F=ma$ becomes
\begin{equation}\label{eq:LLT_forcelaw}
    m \frac{D^2 \psi'(x',t')}{Dt'^2} = k \left[\psi'(x'+\Delta x, t' - \Delta t_+) - 2\psi'(x',t') + \psi'(x'-\Delta x, t' - \Delta t_-) \right] \: .
\end{equation}
 To evaluate this, we perform a Taylor expansion for $\psi'(x' \pm \Delta x,t' - \Delta t_{\pm})$ around the spacetime point $(x',t'_{*} = t' - \Delta t_{0})$ which uses the baseline case of when the fluid particles are not moving (hence simultaneous retarded positions of neighbors on either side, as was the case in medium rest frame $S$), 
    \begin{align}
        \label{eq:Taylorseries2}
    \psi'(x' \pm \Delta x, t' - \Delta t_{\pm}) \: = \: \psi'(x' , t_{*})  \: \pm \: \Delta x \frac{\pdel \psi'}{\pdel x'}\bigg|_{x',t_{*}} \: \mp \: \frac{u \Delta x }{c^{2}} \frac{\pdel \psi'}{\pdel t'}\bigg|_{x',t_{*}} 
    \: + \: \frac{1}{2}\left(\pm \Delta x\right)^{2} \frac{\pdel^{2} \psi'}{\pdel x'^{2}}\bigg|_{x',t_*} \nonumber \\ 
    + \:  \frac{u^2 }{2c^{4}}( \Delta x)^{2} \frac{\pdel^{2} \psi'}{\pdel t'^{2}}\bigg|_{x',t_{*}} 
    \: + \: (\pm \Delta x)\left(\mp \frac{u\Delta x}{c}\right) \frac{\pdel^{2} \psi'}{\pdel x' \pdel t'}\bigg|_{x',t_{*}} \:  + \: \mathcal{O}(\Delta x)^{3} .
    \end{align}
This, when substituted in the force law of Eq. (\ref{eq:LLT_forcelaw}) and neglecting terms of $\mathcal{O}(u/c)^{2}$ and beyond simplifies in the continuum limit $\Delta x \rightarrow 0$ to yield
    \begin{equation}\label{eq:wave_LLT_almost}
        \frac{D^{2}}{Dt'^{2}} \psi'(x',t') \: = \: v^{2} \bigg( \frac{\pdel^{2}}{\pdel x'^{2}}   -\frac{2u}{c^{2}}\frac{\pdel^{2}}{\pdel x' \pdel t'}  \bigg)\psi'(x',t') \: .
    \end{equation}
    We notice explicitly here that $v^2$ arises from the material properties. The combination $v^2/c^2$ (for example, in the mixed derivative term in the above equation) arises from interplay of distinct physics. Thus, $v/c$ is not an expansion parameter. While we neglect terms of order $u^2/c^2$, a term like $uv^2/c^{2}$ is {linear} in $u/c$ and is retained. Terms such as $u^2$ that appear multiplying the second spatial derivative arise already at the level of Galilean kinematics and are therefore retained exactly, as they are not associated with relativistic corrections. Expressing the Lagrangian acceleration in terms of Eulerian derivatives via Eq. (\ref{eq:Lagrangian_acc}), we recover the ``transformed'' form of the wave equation under low-speed limit of Lorentz transformations,
\begin{equation}
    \left[\frac{\partial^2}{\partial t'^2} - (v^2 - u^2)\frac{\partial^2}{\partial x'^2} - 2u \left(1 - \frac{v^2}{c^2} \right) \frac{\partial^2}{\partial x' \partial t'} \right]\psi'(x',t') = 0 \: ,
\end{equation}
which we had found in our earlier analysis of Eq. (\ref{eq:lowlorentz_wave_transformed_linear}) but interpreted here as being the {derived wave equation in an arbitrary inertial frame.} It matches the general factorized invariant form of Eq. (\ref{eq:conjugate_arbit2}) when the left- and right-moving wave components are assigned Lorentz-boosted phase velocities to linear order in $u/c$, that is, \( v_l = v + u - uv^2/c^{2} + \mathcal{O}(u/c)^{2}  \) and \( v_r = v - u + uv/c^{2} + \mathcal{O}(u/c)^{2} \). Finally, as a special case, in the rest frame of the medium, where \( u = 0 \), we recover the canonical wave equation of Eq. (\ref{eq: wave_eq_1D}) again. These findings show that a Newtonian analysis, one that explicitly incorporates retardation effects and uses Lagrangian derivatives, makes the wave equation's invariance manifest in the low-speed relativistic regime, an invariance that was present from the outset in the frame-independence of Newton's second law.

\bigskip
An interesting and non-trivial by-product of this analysis is the definition of a \emph{Lagrangian spatial derivative}, in analogy with the material time derivative of Eq. (\ref{eq:Lagrangian_der}), as (we have adopted unprimed coordinates here to avoid clutter, and we emphasize the Lagrangian spatial derivative is frame‐independent),
\begin{equation}
    \label{Lagrangian_spatial}
    \frac{D}{Dx} = \frac{\partial}{\partial x} - \frac{u}{c^2} \frac{\partial}{\partial t} \: .
\end{equation}
This is because even in the low-speed limit of special relativity, the symmetry between space and time is preserved~\cite{Chen2022}. This operator follows spatial changes as seen by an arbitrary inertial observer, correcting for the relative velocity $u$ and ensuring consistency with the relativity of simultaneity as required by a finite speed of light. In the limit $c \to \infty$, we recover the Galilean result: $D/Dx = \partial/\partial x$, and the wave equation reduces to the one discussed for Galilean relativity of Eq. (\ref{eq:galilean_wave_lagrangian}). The Lagrangian spatial derivative naturally complements the Lagrangian time derivative, forming a coordinate-independent, covariant-like formulation of the wave equation. Together, they yield an invariant form of the wave equation in any arbitrary inertial frame
\begin{equation}
    \label{eq: wave_eq_1D_invariant_Lagrangian}
    \frac{D^2 \psi}{Dt^2} = v^2 \frac{D^2 \psi}{Dx^2} \: ,
\end{equation}
where we recognize the second spatial Lagrangian derivative as the Lagrangian curvature of the wave form in the low-speed (to linear order in $u/c$) relativistic wave equation of Eq. (\ref{eq:wave_LLT_almost})
\begin{equation}
    \label{eq:Lagrangian_curvature}
    \frac{D^{2}}{Dx^{2}} \: = \: \frac{\pdel^{2}}{\pdel x^{2}}   -\frac{2u}{c^{2}}\frac{\pdel^{2}}{\pdel x \pdel t} \: ,
\end{equation}
which accompanies its temporal counterpart of Eq. (\ref{eq:Lagrangian_acc}). More generally, in three dimensions, the Lagrangian spatial derivative $\underline{\vec{\nabla}}$ can be written as,
\begin{equation}
    \underline{\vec{\nabla}} = \vec{\nabla} + \frac{\vec{u}}{c^{2}}\frac{\pdel}{\pdel t} \: ,
\end{equation}
where $\vec{\nabla}$ is the Eulerian gradient and $\vec{u}$ is the velocity of the fluid particle as seen by the observer, giving a fully covariant 3D wave equation,
\begin{equation}
    \frac{D^2 \psi}{Dt^2} = v^2 \: \underline{\vec{\nabla}}^{2}\psi \: .
\end{equation}
It is interesting to note that the parameter $v$ in this covariant low-speed relativistic wave equation depends on parameters of the microscopic interactions and can be interpreted as the wave speed in the rest frame of the medium. In arbitrary frames, the factorized form of Eq. (\ref{eq:conjugate_arbit}) or Eq. (\ref{eq:conjugate_arbit2}) reveals much more transparently the phase velocities of the wave solutions.

\section{Discussion and Outlook}
\label{sec:discussion}
We have shown that the wave equation in any inertial frame can be expressed using Lagrangian derivatives that follow the motion of individual particles
\begin{equation}
    \label{eq: wave_eq_1D_invariant_Lagrangian}
    \frac{D^{2}\psi(x, t)}{Dt^{2}} \: = \: v^{2}  \frac{D^{2}\psi(x, t)}{D x^{2}} \: ,
\end{equation}
where $v = \sqrt{B/\rho}$ is a property of the bulk modulus and density (or generalizations thereof) of the material, and can be interpreted as the wave speed only in the medium rest frame. This wave equation is derived from microscopic force laws for non-relativistic wave phenomenon, compatible with both Galilean and low-speed relativistic frameworks. While the rest frame of the medium appears to ``simplify'' the wave equation by restoring symmetry between left- and right-moving waves, we clarify that it should not be considered a preferred frame in any fundamental sense. The material derivatives $D/Dt$ and $D/Dx$ are defined intrinsically, by following a given material element along its trajectory, and do not themselves refer to any particular frame; the relative velocity $u$ enters only when we re-express them in the Eulerian coordinates of a specific observer. The medium is singled out in just one place, through the material parameter $v = \sqrt{B/\rho}$, which must be defined in the frame where the medium is at rest. This is a physically identifiable frame, distinguished by the presence of the medium, but not a dynamically privileged one: the form of the equation is common to all inertial observers, and only the value of a material constant is tied to the medium frame. In this sense no inertial frame holds a privileged status, even when the physics appears tidier in one frame than another.
Since the microscopic force law $\vec{F} = m\vec{a}$ is valid for small amplitude wave solutions and is itself invariant under both Galilean and low-speed (to linear order in $u/c$) Lorentz transformations, it is expected that the emergent wave equation derived from those force laws should also be invariant. However, this fact is not always made clear in standard pedagogical treatments. The form of the wave equation retains its structure in all inertial frames—only the phase velocities change. By deriving the wave equation from microscopic dynamics using Lagrangian derivatives, we offer a new perspective on a foundational result whose status is sometimes obscured in the literature. We hope this article provides a qualitative and quantitative explanation that helps fill this void. At the same time, it reinforces the argument that Galilean transformation is not the low-speed limit of Lorentz transformation. This, we believe, paired with discussion of wave mechanics can be particularly effective pedagogically.


A physical application of our framework is the Doppler shift. Consider a stationary wave source and a detector moving at speed $u$ relative to the medium. In the Galilean case, one obtains the well-known Doppler-shifted frequency, $\omega' \: = \: \omega\left(1 \pm \frac{u}{v}\right)$, with no change in wavelength $\lambda' = \lambda$ (or equivalently wavenumber $k' = k$). However, when relativistic corrections are taken into account in the low-speed relativistic analysis the situation is more subtle. The wave phase is invariant across frames, $kx - \omega t = k'x' - \omega't',$ and under the low-speed Lorentz transform of Eq. (\ref{eq:lowlorentz}) yields the following transformed frequency and wavenumbers,
\begin{align}
    \omega' &= \omega\left(\frac{1 \pm \frac{u}{v}}{1 - \frac{u^2}{c^2}}\right), \: \: \: \: \: \: \: \: 
    k'= k\left(\frac{1 \pm \frac{uv}{c^2}}{1 - \frac{u^2}{c^2}}\right).
\end{align}
To first order in $u/c$, this recovers the standard frequency Doppler shift while revealing an interesting wavelength shift,
\begin{equation}
    k' \approx k\left(1 \pm \frac{uv}{c^2}\right).
\end{equation}
We emphasize that we do not assume $uv/c^2 \sim \mathcal{O}(1)$; such terms are retained solely because they are linear in the expansion parameter $u/c$. Such wavelength shifts are often neglected in elementary treatments but arise naturally within a first-order relativistic framework. For completeness, we mention that this wavelength shift has been previously discussed in Kong's analysis \cite{kong} of interacting acoustic waves in moving media, an effect we recover through explicit derivation from microscopic dynamics.

Looking ahead, the approach developed here may be extended in several directions. A natural next step is to pursue a fully covariant and relativistic treatment that does not assume small $u/c$, potentially involving modifications to the force law and wave propagation speeds. Additionally, the extension to three spatial dimensions---while technically more involved---would allow exploration of a richer class of wave phenomena. Finally, the pedagogical value of the approach makes it a useful addition to advanced undergraduate instruction, where it can help students unify their understanding of Newtonian dynamics, wave physics, and special relativity.

\section*{Declarations}
H.R.M carried out the initial derivation that was proposed and supervised by A.S. The derivation was then later expanded by A.S. H.R.M wrote the initial draft, which was edited and completed by A.S. The authors have no conflicts to disclose. 

\section*{Acknowledgments}
The authors acknowledge the support of the Fairbanks Research Endowment at Whitman College which helped fund this work. We thank our colleagues, Prof. Moira Gresham and Prof. Doug Juers from Whitman College, and Prof. Suprit Singh from IIT Delhi for helpful discussions, along with the anonymous reviewers, and editors whose comments helped us improve the clarity of the manuscript.

\bibliographystyle{utphys}
\bibliography{MechanicalWave_Relativity}
\end{document}